\documentclass[10pt,a4paper]{article}
\usepackage[utf8x]{inputenc}
\usepackage{ucs}
\usepackage{amsmath}
\usepackage{amsfonts}
\usepackage{amssymb}
\usepackage{amsthm}
\usepackage{algorithm}
\usepackage[noend]{algorithmic}
\usepackage{url}
\author{Wenjie Fang}
\title{A computational approach to the graceful tree conjecture}
\newtheorem{conj}{Conjecture}
\newtheorem{thm}{Theorem}
\begin{document}

\maketitle

\begin{abstract}
Graceful tree conjecture is a well-known open problem in graph theory. Here we present a computational approach to this conjecture. An algorithm for finding graceful labelling for trees is proposed. With this algorithm, we show that every tree with at most 35 vertices allows a graceful labelling, hence we verify that the graceful tree conjecture is correct for trees with at most 35 vertices.
\end{abstract}

\section{Introduction}

Graceful labelling was first introduced by Rosa in \cite{rosa1966certain} under the name "$\beta$-valuation". A graceful labelling of an undirected graph $ G=(V,E) $ is an bijection $ f: V \to \lbrace 0,1,\ldots,Card(V)-1 \rbrace $ such that the function induced $ g: E \to \lbrace 1,2,\ldots,Card(E) \rbrace , \{a, b\} \mapsto |f(a) - f(b)|$ is an injection. In this context, we call $g$ its induced labelling.

A graph which admits a graceful labelling is said to be graceful. After the introduction of this notion of gracefulness, the following conjecture was proposed by Ringel and K\"otzig.

\begin{conj}\emph{(Graceful Tree Conjecture)}
Every tree is graceful.
\end{conj}

It is shown in \cite{rosa1966certain} that this conjecture implies the Ringel's conjecture:

\begin{conj}\emph{(Ringel's Conjecture)}
For all $ n $, for a certain tree $ T $ with $ n $ edges, the complete graph $ K_{2n+1} $ can be decomposed into $ n $ trees, all isomorphic to T.
\end{conj}

According to \cite{gallian2005dynamic}, current approach to the graceful tree conjecture is mostly to prove gracefulness of a certain special kind of trees. Computational approach to verify the conjecture is quite rare. However, as the graceful tree conjecture may be incorrect, efforts for finding a conterexample are also valuable. Furthermore, such verification may provide interesting observation on graceful labeling of trees.

The major result in this direction is from Aldred and McKay \cite{aldred1998graceful}. In their article they verified that every tree with at most $27$ vertices is gracful. They used a stochastic local search algorithm to accomplish this result. Another result is from Horton \cite{horton2003graceful}. In his master dissertation, he claimed a verification of gracefulness for trees with at most $29$ vertices. This result was obtained with a randomized back-tracking algorithm.

Inspired by these results, here we present another computational approach. Using a hybrid algorithm for finding graceful labeling for trees, we obtain the following result.

\begin{thm}
Every tree with at most $35$ vertices is graceful.
\end{thm}

It is achieved by applying a hybrid algorithm to every tree with at most 35 vertices to find a graceful labeling for each of them.

\section{Algorithm}

To enumerate all trees with a certain number of vertices, we use the algorithm proposed in \cite{wright1986constant}, which provides a constant time generation of free trees. Furthermore, we also adopt its representation of trees by level sequences. In this representation, a free tree is rooted in its center if it is a central tree, in one of its bicenters if it is otherwise bicentral. Therefore, every tree can be treated as a rooted tree. In this context, a labelling to vertices can be viewed as a permutation of $\{0, 1, \ldots, Card(V)-1\}$.

The number of vertices is an important factor for performance of algorithms. We note $n = Card(V)$ the number of vertices. As we deal with trees in this article, we know that the number of edges is $n-1$.

The hybrid algorithm consists of two parts, a back-tracking deterministic search and a hill-climbing tabu search combined with some idea from simulated annealing.

\subsection{Deterministic Back-tracking Search}

The back-tracking search is inspired by the master dissertation of Horton \cite{horton2003graceful}. In this dissertation, Horton proposed a randomized back-tracking search for graceful labelling. With his method, he verified that every tree with at most 29 vertices is graceful. Inspired by this method, we proposed a deterministic version of his algorithm with some significant optimization.

Our deterministic search tries to construct a graceful labelling $ f $ with $ f(r)=0 $ for $ r $ the root. This is done by assigning values to vertices one by one. At each recursive call, it tries to make sure that a new value $ k $ appears in the induced labelling $ g $. The value $ k $ decreases as we go deeper in the decision tree, from $ n-1 $ to $ 1 $. This mechanism assures correctness of this algorithm.

To assure that a new value $ k $ appears in the range of $ g $, it finds a not-yet-assigned vertex $ v $ connected to another vertex $ v' $ that is already assigned a label $ f(v') $, then tries to assign to $ v $ a not-yet-assigned label $ f(v) $ such that $ |f(v')-f(v)|=k $. There may be several possibilities, or none. If this attempt fails, it tracks back, restores its status and pursues another possiblity if there exists one.

As the decision tree can grow exponentially in size when $n$ increases, we manually add a threshold on the number of backtracks. This prevents searching for a very long time. This threshold is tuned with respect to the performance of the probabilistic search described. It is empirically fixed to $ (n-19)*11000 - 1000 $ in this verification. For a new, improved version of probabilistic search, it is empirically fixed to $ (n-18)*1000 $.

Here is a pseudo-code of the deterministic back-tracking search.

\algsetup{indent=2em}
\begin{algorithm}
\caption{Back-tracking Search}
\begin{algorithmic}
\REQUIRE A new value $ k $ not yet appeared in the induced labelling $ g $
\ENSURE Return a boolean indicating whether the search is sucessful
\STATE \textbf{Function} Search($k$)
\IF{$ k=0 $}
\RETURN \TRUE \COMMENT{This indicates that every vertice is properly labeled.}
\ENDIF
\IF{iterations exceed threshold} 
\RETURN \FALSE 
\ENDIF
\FOR {every vertex $ v $ without label, with its parent $ v' $ labeled}
\IF{label $ f(v') + k $ valid and not yet used}
\STATE $ f(v) \gets f(v') + k $, update tables
\IF{Search($ k-1 $)} 
\RETURN \TRUE 
\ENDIF
\STATE Restore tables, unassign $ v $
\ENDIF
\IF{label $ f(v') - k $ valid and not yet used}
\STATE $ f(v) \gets f(v') - k $, update tables
\IF{Search($ k-1 $)}
\RETURN \TRUE
\ENDIF
\STATE Restore tables, unassign $ v $
\ENDIF
\ENDFOR
\RETURN \FALSE
\end{algorithmic}
\end{algorithm}

For efficiency, this algorithm maintains a table of already assigned labels and a linked list of vertices not yet assigned but connected to a vertice already with a label. 

Another optimization is also used in the maintenance of this linked list. It is obvious that, if a vertex has two children that are isomorphic, i.e. they induce isomorphic subtrees, we can permute labels of these subtrees. Using this symmetry, we can eliminate redondant isomorphic vertices by a precomputation. This trick is from \cite{horton2003graceful}.

For choosing vertex to label, there are two natural ways, all relying on the representation of level sequences. Each item in the representation represents a vertex. One way is to choose from left to right, another from right to left. Their performance is different. Right-to-left strategy is more efficient, probably because it reduce the problem's size by assigning trivial labels to leaves attached to the root.

To initialize, we simply initialize every table, assign label $ 0 $ to the root, then start to do back-tracking. 

Though efficient, this back-tracking search misses a lot of potential solutions, due to the choice to label the root $ 0 $. For a more thorough search, another algorithm is needed.

\subsection{Probabilistic Search}
This probalilistic search is inspired by Aldred, McKay\cite{aldred1998graceful}. They proposed a searching method for graceful labeling which is a combination of tabu search and hill-climbing. With that method, they verified that every tree with at most 27 vertices is graceful. Our probabilistic search is a hill-climbing tabu search combined with some ideas from simulated annealing.

This approach lies in the field of combinatorial optimization, whose goal is to find an extremum of a certain evaluation function in a discrete domain. In an attempt to solve a decision problem, like determining whether a tree is graceful, by combinatorial optimization, it is a common practice to propose a evaluation function whose minimum reaches a certain value if and only if the answer to the corresponding decision problem is positive. Therefore, by finding the minimum of this particular evaluation function, we can determine the answer of the decision problem.

For determining whether a tree is graceful, it is natural to search for potential graceful labelling, which can be viewed as permutations.

In verifications mentioned in this article, the following evaluation function is used. It is always positive, and only reaches $0$ when $f$ is a graceful labelling.
\[ Eval(f) = \sum_{i \in \lbrace 1, \ldots, n-1 \rbrace \setminus Im(g)} i, \] 
where 
\[ Im(g)=\lbrace |f(x)-f(y)| \big| \lbrace x,y \rbrace \in E \rbrace \]

However, we discovered the following evaluation function which is more efficient.
\[ Eval'(f) = \sum_{i \in \lbrace 1, \ldots, n-1 \rbrace \setminus Im(g)} i, \] 
where 
\[ Im(g)=\lbrace |f(x)-f(y)| \big| \lbrace x,y \rbrace \in E \rbrace \]

$ Eval'(f)=0 $ if and only if the function induced $ g $ is injective, which is equivalent to the gracefulness of the current tree with $ f $ its graceful labelling. Furthermore, $ Eval(f) $ is always positive. This evaluation function also stresses differences between values in the domain $ \lbrace 1, \ldots, n-1 \rbrace $. For $ k \in \lbrace 1, \ldots, n-1 \rbrace $, there are $ k $ combinations of $ \lbrace f(x),f(y) \rbrace $ that have a difference $ |f(x)-f(y)| = n-k $. As a result, penalties on lack of large difference should be more severe than small difference, as there are fewer combinaisons for large difference. Our evaluation function reflects exactly this fact.

Here is a pseudo-code of our probabilistic search. We consider here a potential graceful labeling as a permutation of $ \lbrace 0, \ldots, n-1 \rbrace $.

\algsetup{indent=2em}
\begin{algorithm}
\caption{Probabilistic search using metaheuristics}
\begin{algorithmic}
\REQUIRE A labeling $ f $ being a permutation
\ENSURE Return a graceful labeling if one exists, loop otherwise
\STATE $ v \gets Eval(f) $ where $ Eval $ is an evaluation function
\WHILE{$ v \neq 0 $}
\STATE Randomly choose $ 2n $ pairs of vertices
\FORALL{pair of vertices $ {x,y} $ chosen}
\STATE Swap $ f(x),f(y) $
\STATE Calculate $ Eval $ for the modified labeling
\STATE Restore status
\ENDFOR
\STATE Choose the pair $ {x,y} $ that minimize $ Eval $
\STATE $ v' \gets Eval(f') $ where $ f' $ is obtained by swapping $ f(x),f(y) $ in $ f $
\IF{$ f(x),f(y) $ was not swapped in previous $ \lfloor n/3 \rfloor $ iterations}
\IF{$ v > v' $}
\STATE Swap $ f(x),f(y) $, update $ v $ by $ v \gets v' $
\ELSE
\STATE With a probability $ p $, swap $ f(x),f(y) $, update $ v $
\ENDIF
\ELSE
\IF{$ v=0 $}
\STATE Swap $ f(x),f(y) $, update $ v $
\ENDIF
\ENDIF
\STATE Update the table storing previous swaps
\ENDWHILE
\RETURN f
\end{algorithmic}
\end{algorithm}

By minimizing $ Eval $, we can efficiently explore labellings that are likely to be graceful. We use hill-climbing (or in this case, hill-descending) to try to minimize $ Eval $. At each iteration, the algorithm tries a number of random modifications and picks the one with best evaluation. This number is fixed emprically to $ 2n $.

It is known that hill-climbing method can be trapped in a local minimum. In order to avoid this problem, we use a tabu search. The algorithm keeps track of a number of previous modifications and forbids such modification unless the result is a graceful labeling. Therefore, the algorithm always go forth to search for new solutions. We fix the number of forbidden previous modifications to $ \lfloor n/3 \rfloor $. This value is determined empirically.

Also proposed to solve the local minimum problem, the algorithm accepts, with a certain probability determined empirically, modifications that worsen the solution. This behavior is intended to emulate simulated annealing, which can escape local minimum with a probability determined by its ``temperature''.

To initialize, we provide to the algorithm an identity permutation. In practice, as the algorithm do not guarantee termination, a threshold is enforced on the number of iterations. Trees that make this algorithm reaches threshold are recorded and examined afterwards.

This algorithm does not guarantee to produce a graceful labeling. Theoretically, it can go into an infinity loop. But in practice, this has never happened. Furthermore, meta-heuristics employed in this algorithm generally have good chances to avoid looping and escaping from local minimum. We are optimistic that this algorithm will eventually return a graceful labeling if there exists one, though it is not proved.

\subsection{Hybrid algorithm}

Combining this two algorithm, we obtain a hybrid algorithm for finding a graceful labeling for a tree. 

There is two stages in the hybrid algorithm. In the first stage it applies the deterministic back-tracking search. If it fails to find a graceful labeling, it turns to the second stage, where it performs the probabilistic search.

The reason of this strategy is as following. The deterministic back-tracking outperforms the probabilistic search in most cases, but in some cases it will take an enormous amount of time. The probabilistic search is not fast comparing to the deterministic one, but its runtime varies in a much more moderated way. Therefore, it is natural to use the deterministic search with a cutoff of runtime, then patch unfinished cases with the probabilistic search.

Experimental results show that this hybrid approach outperforms both the deterministic and the probabilistic search algorithm.

\section{Results and observations}

By applying the hybrid algorithm to every tree with at most $ 35 $ vertices, we verify that every such tree is graceful.

Below shows various statistics on this hybrid algorithm. Runtime data are collected on an Intel Core 2 Duo T7200.

\medskip 

\begin{tabular}{|c|r|r|r|l|c|c|}
\hline
n & trees & backtracking & remainder & ratio & time(s) & avg(ms)\\
\hline 
20 & 823065 & 823002 & 63 & 7.65e-5 & 11 & 0.0134 \\
\hline
21 & 2144505 & 2144461 & 44 & 2.05e-5 & 32 & 0.0149 \\
\hline
22 & 5623756 & 5623588 & 168 & 2.99e-5 & 106 & 0.0188 \\
\hline 
23 & 14828074 & 14827895 & 179 & 1.21e-5 & 325 & 0.0219 \\
\hline
24 & 39299897 & 39298893 & 1004 & 2.55e-5 & 1041 & 0.0265 \\
\hline
25 & 104636890 & 104635672 & 1218 & 1.16e-5 & 3098 & 0.0296 \\
\hline
26 & 279793450 & 279787959 & 5491 & 1.96e-5 & 9800 & 0.0350 \\
\hline
27 & 751065460 & 751056670 & 8790 & 1.17e-5 & 29450 & 0.0392 \\
\hline
28 & 2023443032 & 2023410238 & 32794 & 1.62e-5 & 92218 & 0.0456 \\
\hline
29 & 5469566585 & 5469504091 & 62494 & 1.14e-5 & 279845 & 0.0512 \\
\hline
30 & 14830871802 & 14830672030 & 199772 & 1.35e-5 & 864580 & 0.0583 \\
\hline
31 & \multicolumn{6}{|c|}{*}\\
\hline
32 & \multicolumn{6}{|c|}{*}\\
\hline
33 & 300628862480 & 300625170528 & 3691952 & 1.23e-5 & * & * \\
\hline
34 & 823779631721 & 823768359223 & 11272498 & 1.37e-5 & * & * \\
\hline
35 & 2262366343746 & 2262333140305 & 33203441 & 1.47e-5 & * & *\\
\hline
\end{tabular}

\medskip 

The second column indicates the total number of trees with $ n $ vertices. The third column indicates the number of trees proved to be graceful with the back-tracking search. The fourth column is the number of trees proved to be graceful by the probabilistic search. The fifth column indicates the proportion of trees failing to be proved graceful in the back-tracking search stage. The sixth column indicates the calculation time needed in second. The seventh indicates average calculation time for each tree in millisecond.

Verification of gracefulness for trees with $ 31 $ or $ 32 $ vertices is done with an older version of our algorithm. As parameters differ, data are omitted for consistency. Verification of gracefulness for trees with $ 33 $, $ 34 $ or $ 35 $ vertices is accomplished with the help of a Chinese community of volunteer computing. The whole task is divided into small fragments and done on heterogeneous machines. Therefore, no accurate timing can be provided. This volunteer computing effort is organized with a website now located at \url{http://www.eleves.ens.fr/home/wfang/gtv/index_en.html}.

Given the statistic, we can see that the back-tracking algorithm is extremely efficient. It can find a graceful labeling for nearly every tree (more than $ 99.99\% $) in a short time (at most $0.1$ms for each tree). Though encouraging, this may only be an illusion of limited data, as it lacks support of sufficient data for large $ n $. On the other hand, it seems that this efficiency can be extrapolated for a few points further. 

Regarding calculation time, by regression, we can show the relations below. ($ T $ for total calculation time in second, $ T_{avg} $ for calculation time for each tree in second)

\[
T \approx O(3.09^{n}), T_{avg} \approx O(1.16^{n})
\]

By extrapolation, verification of gracefulness for trees with $ 35 $ vertices should take $ 7.7 $ years on a core of Core 2 Duo T7200. We should note that this relation is only valid for current tests. The validity of extrapolation is not guaranteed.

Though runtime seems to grow exponentially with the number of vertices, the time to find a graceful labeling for one tree increases rather slowly. Hence it is reasonable that the performance would stay to be good for small values of $ n $, which make this hybrid algorithm seems suitable for further verification of the graceful tree conjecture.

\section{Acknowledgement}
Thanks to volunteers from Team China, without whom this verification is impossible. Thanks for their donation of computational power, their support and their enthusiasm.

\bibliography{graceful}{}
\bibliographystyle{plain}
\end{document}